\documentclass[useAMS,usenatbib,nofootinbib,onecolumn]{mn2e}
\usepackage{lmodern}

\usepackage[T1]{fontenc}
\usepackage[utf8]{inputenc}
\usepackage{amsmath}
\usepackage{amssymb}
\usepackage{graphicx}
\usepackage{esint}

\usepackage{hyperref}
\usepackage[usenames,dvipsnames]{color}
\definecolor{menublue}{rgb}{0.0,0.0,0.5}
\definecolor{citegreen}{rgb}{0.0,0.5,0.5}
\definecolor{urlred}{rgb}{1.0,0.0,0.0}
\hypersetup{bookmarksopen,pdfstartview={FitH},colorlinks=true, breaklinks=true,menucolor=menublue,urlcolor=urlred,citecolor=citegreen,linkcolor=blue}
\usepackage[all]{hypcap}

\makeatletter

\pdfpageheight\paperheight
\pdfpagewidth\paperwidth

\renewcommand{\[}{\begin{equation}}
\renewcommand{\]}{\end{equation}} 

\makeatother

\title[A new tool for Fisher matrix forecasts]{Optimizing parameter constraints: a new tool for Fisher matrix forecasts }

\author[L. Amendola, E. Sellentin]
{Luca Amendola$^1$ and Elena Sellentin$^1$\\
$^1$Institut f{\"ur} Theoretische Physik, Universit{\"a}t Heidelberg, Philosophenweg 16, 69120 Heidelberg, Germany}

\begin{document}
\maketitle

\begin{abstract}
In a Bayesian context, theoretical parameters are correlated random
variables. Then, the constraints on one parameter can be improved
by either measuring this parameter more precisely - or by measuring
the \emph{other} parameters more precisely. Especially in the case
of many parameters, a lengthy process of guesswork is then needed
to determine the most efficient way to improve one parameter's constraints.
In this short article, we highlight an extremely simple analytical
expression that replaces the guesswork and that facilitates a deeper
understanding of optimization with interdependent parameters. 
\end{abstract}

\date{\today}


\section{Introduction}

The Fisher matrix has become widely spread in cosmology since it allows
quick forecasts of parameter constraints in the limit of a Gaussian
posterior likelihood. Recent works on the Fisher matrix extend the
formalism also to errors in the independent variables \citep{GenFish}
and to non-Gaussianity \citep{DALI1,DALI2,Joachimi}. Its ease of
handling can be traced back to many analytical operations that can
be performed on a multivariate Gaussian and that can be summarized
in a manual-like collection: 
\begin{itemize}
\item in order to maximize a parameter, i.e. fixing it to its bestfit value, the rows and columns of this parameter
must be removed from the Fisher matrix 
\item the removal of rows and columns from the inverse of the Fisher matrix
leads to a marginalization of the respective parameter 
\item a combination of independent experiments with the same fiducial or
best fit can be achieved by adding up their Fisher matrices 
\item a transformation of variables can be achieved by multiplying the Fisher
matrix on the left and the right with the Jacobian matrix of the transformation and its
transpose. 
\end{itemize}
To this Fisher matrix manual, we now want to add a further rule that
allows optimizing an experiment, answering the question: Given $n$
correlated parameters $\theta_{1},...\theta_{n}$, and not being able
to improve the measurement of the parameter $\theta_{i}$, which \emph{other}
parameter $\theta_{k}$ should we measure more precisely in order
to best improve the constraints on $\theta_{i}$, and which gain in
the precision of $\theta_{i}$ can then be expected? For clarity,
in the following we refer to the $i$-th parameter as the \emph{target}
parameter and to the $k$-th parameter as the \emph{control} parameter which shall emphasise that we expect to better control the accuracy of this parameter with future experiments.

This problem can occur in a variety of situations. Let us take a typical
case in cosmological applications. In a standard analysis of CMB data
(see e.g. \citet{PlanckXV}), one has parameters that depend on early
cosmology (e.g. the spectral index $n_{s}$), parameters that depend
on the local late-time universe, e.g. the Hubble constant $H_{0}$, and parameters
that depend on totally different physics, e.g. the ``nuisance''
parameters that describe some foreground contamination, e.g. the amplitude
of the cosmic infrared background $A^{cib}$. The experiments or the
theoretical arguments that can put additional constraints on, say,
$A^{cib}$ can be completely different from those that measure $n_{s}$
and a sensible question to ask is how much one can improve the estimation
of $n_{s}$ (our target parameter) when adding more constraints on
$A^{cib}$ (the control parameter). The usual way to answer this question
is to invert the parameter covariance matrix (which is normally what
the experiment provides) deriving a Fisher matrix, add priors to the
Fisher matrix that shall quantify the expected future gain in measurement precision of the control parameters, and invert again. The updated covariance of the target
parameter will then be a diagonal element of this inverse matrix. If the prior is found to produce a strong decrease in the variance of the target parameter, one would typically undertake even big scientific efforts in order to really establish a measurement that can produce such a constraint as the prior. On the other hand, if the target parameter does not react sensitively to an improvement in the constraint of the control parameter, one would not need to measure the control parameter better. 
 
Searching for effective priors by adding them to a Fisher matrix and inverting is \emph{per se} a simple operation, but repeating it for any
pair of target/control parameters and for any possible value of the
prior will rapidly become a tedious and lenghty process if the covariance
matrix contains dozens of parameters. Additionally repetitive inversions
are prone to numerical uncertainties. The complexity of possible combinations
of target/control parameters further increases if a correlated \emph{set}
of control parameters shall be improved upon with priors. The rest
of this short note is devoted to deriving and discussing an analytical
formula that drastically simplifies the procedure. 

At the time of the first version on this archive, we wrote that "To the best of
our knowledge, this simple formula seems not to have been pointed
out before". After publishing this paper, however, Eric Linder made us notice that the formula has actually been derived earlier in \citet{Astier:2000as}.

\section{The Sherman-Morrison-Woodbury formula}
\begin{figure}
\begin{center}
\includegraphics[width=0.45\textwidth]{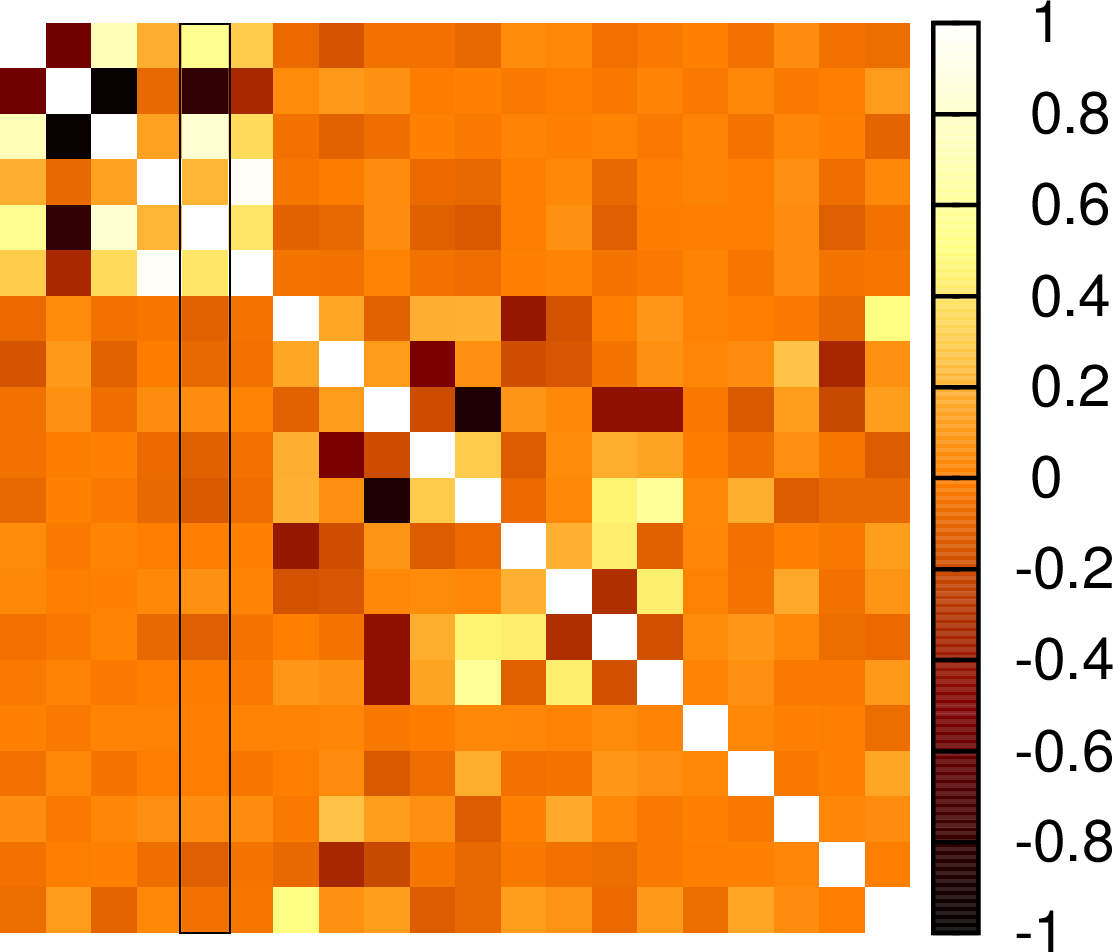} \protect\caption{Correlation matrix of the Planck cosmology and nuisance parameters.
The first 6 columns and rows are the cosmological parameters $\omega_{b},\omega_{cdm},H_{0}$,$A_{s}$,$n_{s}$,$\tau_{reio}$ which are strongly correlated amongst each other. The remaining columns are the nuisance parameters $A^{ps}_{100}$,$A^{ps}_{143}$,$A^{ps}_{217}$,$A^{cib}_{143}$,$A^{cib}_{217}$,$A_{sz},\rho_{ps},\rho_{cib}$,
$n^{Dl}_{cib}$,$cal_{100}$,$cal_{217}$,$\xi^{sz}_{cib}$,$A_{ksz}$,$Bm_{1,1}$ whose meanings are explained in \citep{PlanckXVI}. The marked column contains the correlations of $n_{s}$.}
\label{Cormat} 
\end{center}
\end{figure}

The idea is entirely based on the Sherman-Morrison-Woodbury formula \citep{sherman1950, Woodbury}.
This formula states that if $\sf{M}$ is a square matrix and $\boldsymbol{u},\boldsymbol{v}$ are
vectors then 
\begin{equation}
(\sf{M}+\boldsymbol{uv}^{T})^{-1}=\sf{M}^{-1}-\frac{\sf{M}^{-1}\boldsymbol{uv}^{T}\sf{M}^{-1}}{1+\boldsymbol{v}^{T}\sf{M}^{-1}\boldsymbol{u}}
\label{Sh}
\end{equation}
where $T$ denotes transposition. The formula allows to quickly find
the inverse of $\sf{M}$ when a matrix $\boldsymbol{uv}^{T}$ is added to $\sf{M}$. If we
demand that $\sf{M}$ is a non-degenerate Fisher matrix, i.e. the inverse
of the covariance matrix of parameters, then $\sf{M}$ is symmetric and
positive definite. If we further set $\boldsymbol{u}=\boldsymbol{v}$, we can construct a prior
matrix $\sf{P}=\boldsymbol{uu}^{T}$ where the vector $\boldsymbol{u}$ is 
\begin{equation}
\boldsymbol{u}=\{0,0,...p_{k}^{-1},...\}=p_{k}^{-1}\hat{\boldsymbol{u}}
\end{equation}
where $p_{k}$ is the prior standard deviation on the control parameter $\theta_k$, that shall quantify the expected improvement on $\theta_k$ in a feasible future experiment. The vector $\boldsymbol{\hat{u}}$ is the $k$-th basis vector. Introducing a matrix $\hat{\sf{P}}=\hat{\boldsymbol{u}}\hat{\boldsymbol{u}}^{T}$ i.e. a matrix whose elements are all zero except for $\hat{\sf{P}}_{kk} = 1$, we can write ${\sf{P}}=p_k^{-2}\hat{\sf{P}}$, and  Eq.~(\ref{Sh}) then specializes to 
\begin{equation}
(\sf{M}+\sf{P})^{-1}=\sf{M}^{-1}-\frac{\sf{M}^{-1}\sf{PM}^{-1}}{1+\mathrm{Tr}(\sf{PM}^{-1})}.
\end{equation}
The variance $\sigma_i^2$ of our target parameter $\theta_i$, after having marginalized over all other parameters, is given by $\sigma_{i}^{2}=\sf{M}^{-1}_{ii}$. Its improved
variance $\sigma_{i,\rm{new}}^{2}$ after
having added a prior to the $k$-th control parameter is then 
\begin{align}
\sigma_{i,\rm{new}}^{2} & =({\sf{M+P}})_{ii}^{-1}\\
& = {\sf{M}}_{ii}^{-1}-\frac{({\sf{M}}^{-1}{\sf{PM}}^{-1})_{ii}}{1+\mathrm{Tr}({\sf PM}^{-1})}\\
& = {\sf{M}}_{ii}^{-1}-\frac{p_{k}^{-2}({\sf{M}}^{-1}\hat{{\sf{P}}}{\sf{M}}^{-1})_{ii}}{1+p_{k}^{-2}\mathrm{Tr}(\hat{{\sf{P}}}{\sf{M}}^{-1})}\\
 & = \sigma_{i}^{2}-\frac{p_{k}^{-2}({\sf M}^{-1}\hat{{\sf P}}{\sf M}^{-1})_{ii}}{1+p_{k}^{-2}\sigma_{k}^{2}}\\
& = \sigma_{i}^{2}-\frac{({\sf M}^{-1}\hat{{\sf P}}{\sf M}^{-1})_{ii}}{p_{k}^{2}+\sigma_{k}^{2}}
\end{align}
Now we use the fact that the fully marginalized variance of the control parameter is $\mathrm{Tr}(\hat{{\sf P}}{\sf M}^{-1})=\sigma_{k}^{2}$
and 
\begin{equation}
({\sf M}^{-1}\hat{{\sf P}}{\sf M}^{-1})_{ii}=\rho_{ik}^{2}\sigma_{i}^{2}\sigma_{k}^{2}
\end{equation}
where $\rho_{ik}={\sf M}_{ik}^{-1}/\sqrt{{\sf M}_{ii}^{-1}{\sf M}_{kk}^{-1}}$ is the
correlation coefficient and one has $|\rho_{ik}|\le1$ because of positive-definiteness
of ${\sf M}$. So we derive our main result 
\begin{align}
\sigma_{i, \rm{new}}^{2} & =\sigma_{i}^{2}-\frac{\rho_{ik}^{2}\sigma_{i}^{2}\sigma_{k}^{2}}{p_{k}^{2}+\sigma_{k}^{2}},\label{MainRes}
\end{align}
which describes directly and transparently how the variance of the target parameter $\theta_i$ decreases if we measure better the parameter $\theta_k$.

Also, if the prior on the control parameter $k$ is very weak, i.e.
$p_{k}\to\infty$, the error  $\sigma_{i}$ of the target parameter,
does not change. This equation can be trivially applied even when
the control parameter coincides with the target parameter, by putting
$i=k$ and the self correlation $\rho_{ik}=1$.

From the previous equation, the decrease $\Delta \sigma_{i}^{2} = \sigma_{i,\rm{new}}^2 - \sigma_i^2 $ follows to be
\begin{equation}
\frac{\Delta\sigma_{i}^{2}}{\sigma_{i}^{2}}=-\frac{\rho_{ik}^{2}}{1+\varepsilon}\label{eq:main}
\end{equation}
where $\varepsilon=p_{k}^{2}/\sigma_{k}^{2}$. This tells us that
if we add a prior to the error on the control parameter which is $\varepsilon$
times the current error, then the target parameter constraint decreases
by a fraction $\rho_{ik}^{2}/(1+\varepsilon$). At most, the fractional
decrease is then $\rho_{ik}^{2}\le1$. So the very simple recipe for
choosing the most convenient control parameter to improve the estimation
of the target parameter, is to select the most correlated one. This
of course was to be entirely expected; our formula (\ref{eq:main})
quantifies the effect in a very simple way as a function of the correlation
coefficient and of the ratio $\varepsilon$.

A generalization to several control or target parameters is described
in the Appendix.

\section{Example: Indirectly improving CMB constraints on the primordial slope}

\begin{figure}
\begin{center}
\includegraphics[width=0.45\textwidth]{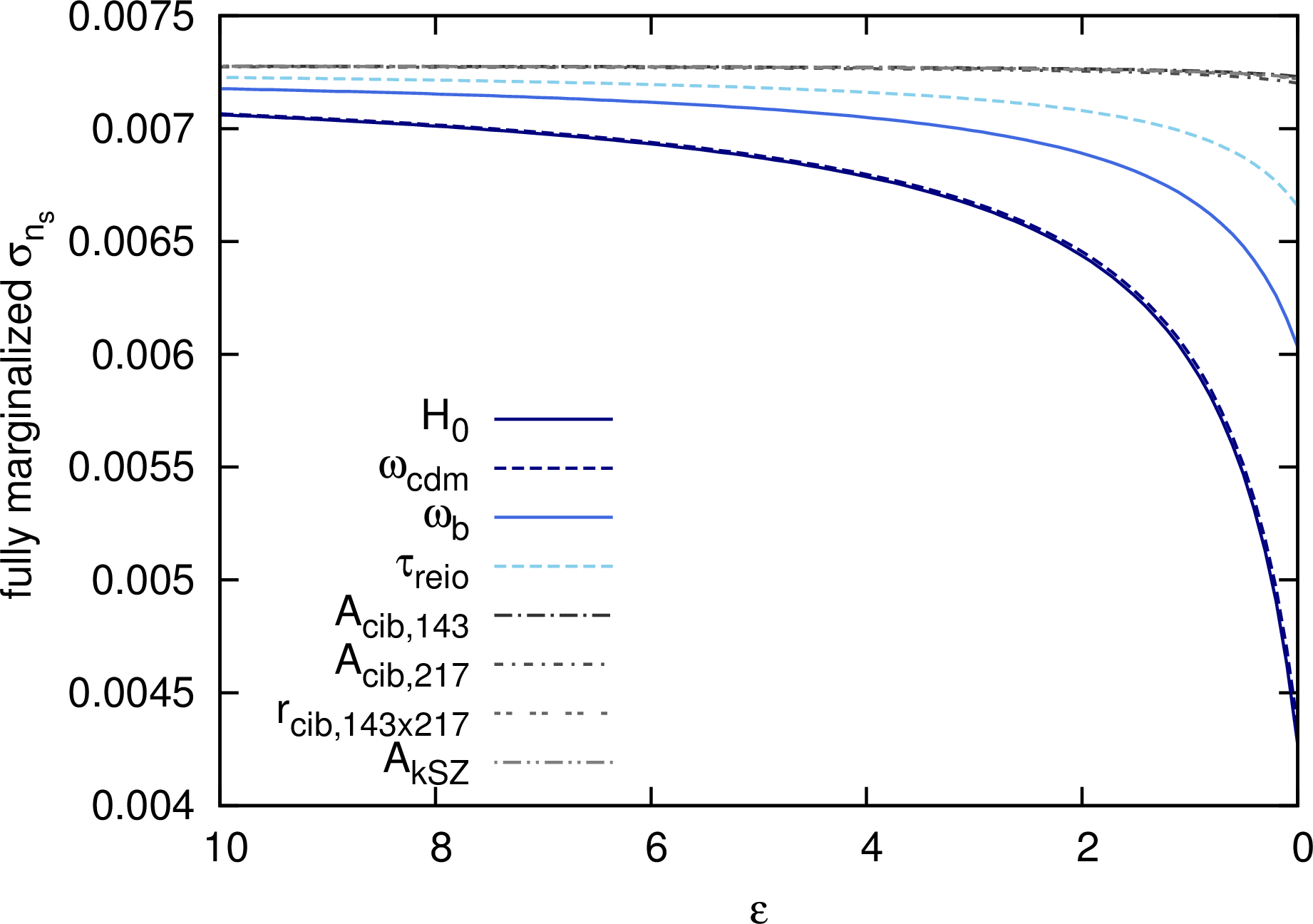} \protect\caption{Improvement on the fully marginalized standard deviation of $n_{s}$,
when adding a prior $p_{k}^{2}=\epsilon\sigma_{k}^{2}$ on the $k$th
control parameter (as indicated by the legend). For $\epsilon=0$,
the $k$th control parameter is exactly known, and the remaining error
on $n_{s}$ stems from the marginalization of other parameters and
the measurement noise on $n_{s}$ itself. }
\label{Shermi} 
\end{center}
\end{figure}
For a mere illustrative purpose, we consider now the actual parameter
covariance matrix $\sf{C}$ for Planck \citep{PlanckXV} as published in
the package Monte Python \citep{MontePython} and assume that the
posterior is well approximated by a multivariate Gaussian, such that we can interpret $\sf{C}$ as an inverse Fisher matrix.

Since Planck does not take any data anymore, it will not improve directly the measurement of any cosmological parameters. It is however well known that Planck has many nuisance
parameters that are degenerate with the primary cosmological parameters
\citep{PlanckXVI}. Therefore, the constraining power of the Planck data set can be boosted by constraining these nuisance parameters better, but how much gain can be expected from this, and which nuisance parameter should be tackled with the most effort?
For example, the amplitude of the cosmic infrared
background (CIB) is a nuisance parameter for Planck, and can in principle
be constrained with new observations independent of Planck. The same
goes for other cosmological parameters, e.g. the Hubble constant $H_{0}$. 

Let's assume we want to constrain the spectral index $n_{s}$ more
precisely because we are interested in inflationary physics, i.e. $n_{s}$ is our target parameter.
With the formalism developed in the previous section, we can now search for the most
effective control parameter to achieve this: We use the Planck baseline
covariance matrix ${\sf C}$ and dissect it into a correlation matrix ${\sf R}$
\begin{equation}
{\sf R}_{ij}=\frac{{\sf C}_{ij}}{\sqrt{{\sf C}_{ii}}\sqrt{{\sf C}_{jj}}}
\label{eq:correl}
\end{equation}

The resulting correlation matrix is depicted in Fig.~(\ref{Cormat}).
From the fifth column we see that $n_{s}$ is most correlated with
the Hubble constant $H_{0}$ with a correlation coefficient of $\rho_{n_{s},H_{0}}=0.81$.
So constraining $n_{s}$ indirectly can best be achieved by better
constraining the Hubble constant $H_{0}$ through local measurements,
see e.g. \citet{3percHubble}.

Second best choice would be constraining $\omega_{cdm}$ ($\rho_{n_{s},\omega_{cdm}}=-0.8$),
third best would be constraining $\omega_{b}$, due to $\rho_{n_{s},\omega_{b}}=0.56$.
Next follows the optical depth $\tau_{reio}$ with $\rho_{n_{s},\tau_{reio}=0.4}$
which can be independently constrained through radio observations
of the dark ages. The best choices for improving upon nuisance parameters
are the parameters that specify Planck's phenomenological model of
the cosmic infrared background. Planck uses a power-law model
for the CIB-spectrum, whose amplitudes in the different Planck channels
are parameterized by the nuisance parameters $A_{143}^{cib}$, $A_{217}^{cib}$
and the correlation between the two channels is given by the nuisance
parameter $\rho_{143\times217}^{cib}$. The kinetic Sunyaev-Zeldovich
effect also possesses a relatively high correlation with $n_{s}$.
All of these nuisance parameters are correlated with $n_{s}$ on the
order of $0.1$ so that we can expect at most an improvement of 1\%
on $n_{s}$ for each perfectly measured nuisance parameter.

In Fig.(\ref{Shermi}), employing Eq. (\ref{MainRes}), we plot by
how much the constraints on $n_{s}$ reduce, given an improved estimate
of any of the named parameters. For instance, an independent prior
on $H_{0}$ that is as good as the current Planck uncertainty (i.e.
$\varepsilon=1$) will decrease the variance of $n_{s}$ by $0.81^{2}/2$
times the old variance, i.e. $\sigma_{n_{s}}$ will improve by 18\%.
At most, the error on $n_{s}$ can improve by 41\% by a perfect determination
of $H_{0}$. By applying the Eq. (\ref{eq:detratio-1}) in Appendix,
we obtain a reduction of the standard deviation from the current Planck value $\sigma_{n_s} = 0.0073$ down to $\sigma_{n_{s}}=0.0063$
if all the nuisance parameters were precisely known and $\sigma_{n_{s}}=0.0025$
if all the parameters, except obviously $n_{s}$ itself, were precisely known.

\section{Conclusions}
The Sherman-Morrison-Woodbury formula allows to calculate the inverse of a
perturbed matrix. We have described its application to a covariance matrix or a Fisher matrix, whose
inverse is of particular interest since its diagonal elements represent
the fully marginalized errors on parameters. We used priors in order to parameterize the gain by potential future experiments or theoretical efforts, that would only be undertaken if they effectively break degeneracies and thereby lead to big improvements in the fully marginalized errors.

Under the addition of
priors to the Fisher matrix, the marginalized errors of course decrease. However,
the magnitude of the decrease could not be foreseen in a transparent way by the standard
procedure of inverting the Fisher matrix. The Sherman-Morrison-Woodbury formula instead provides an \emph{analytical} result that allows to judge quickly
how the addition of priors on one parameter, will propagate through to the constraints
on other parameters. 

The conclusion of this paper is summarized in
a new bullet point to the manual of the Fisher matrix: 
\begin{itemize}
\item When a prior $p_{k}$ on parameter $\theta_{k}$ is added, then the
variance of the parameter $\theta_{i}$ will decrease as prescribed
by Eq.~(\ref{MainRes}). 
\end{itemize}
This new rule provides an effective guidance to where modern cosmology should put the most effort, in order to quickly break parameter degeneracies and improve the constraints of interdependent parameters. For an exactly (approximately) Gaussian posterior likelihood, this result holds exactly (approximately). If the posterior is moderately non-Gaussian, the covariance matrix (as estimated from an MCMC run) and the Fisher matrix, tend to not agree anymore. Applying the here presented tool to the MCMC-covariance matrix might then still provide a good guidance, although it should not be applied to the Fisher matrix. If however, there exists additionally a strongly pronounced genuine $n$-point function for $n > 2$, then analytical marginalizations are not possible anymore, and the analytical tool here presented is no longer well suited. Then, refuge to numerical brute force must be sought.

\section*{Acknowledgments}
We thank Eric Linder for pointing out the earlier derivation of Eq. \ref{MainRes}. We also acknowledge support from DFG throughout the project TRR33 "The Dark Universe".

\section*{Appendix: generalization to several parameters}

Eq. (\ref{MainRes}) can be generalized to several control parameters
but the explicit formulae for the general case become rapidly very
cumbersome. Any symmetric prior matrix ${\sf P}$ acting on $N$ control
parameters (i.e. containing non-zero entries only for a subset of
$N$ rows and columns) can be written as 
\begin{equation}
{\sf P} =\sum_{k=1}^{N}\lambda_{k}\boldsymbol{u}_{k}\boldsymbol{u}_{k}^{T}\equiv\sum_{k=1}^{N}{\sf P}_{k}
\end{equation}
where $\lambda_{k}$ are the eigenvalues of ${\sf P}$ and $\boldsymbol{u}_{k}$ is the
$k$-th orthonormal eigenvector. Then the new Fisher matrix becomes
${\sf M}+\sum_{k}{\sf P}_{k}$ where every ${\sf P}_{k}$ is in the form required for
Eq. (\ref{Sh}) and repeated application of the Sherman–Morrison–Woodbury formula
generates the result. For just two control parameters, $\theta_n$ and $\theta_k$,
the final result will depend on the three independent entries of the
symmetric submatrix ${\sf M}_{nk}$, the three independent entries of the
prior matrix that add information on $\theta_n$ and $\theta_k$, the two correlation coefficients
$\rho_{in},\rho_{ik},$ and on the initial variance $\sigma_{i}^{2}$, so a total of 9 quantities. For $N$ control parameters the formula
depends on $(N+1)^{2}$ quantities.

A prior matrix ${\sf P}$ that adds information on the
two control parameters $\theta_n$ and $\theta_k$ will only have three distinct
non-zero elements ${\sf P}_{kk},{\sf P}_{nn}$ and ${\sf P}_{nk}$. It can be decomposed
as 
\begin{equation}{\sf P}=\lambda_{1}\boldsymbol{u}_{1}\boldsymbol{u}_{1}^{T}+\lambda_{2}\boldsymbol{u}_{2}\boldsymbol{u}_{2}^{T}
\end{equation}
 where,
$\boldsymbol{u}_{1}=(\cos\phi,-\sin\phi)$ and $\boldsymbol{u}_{2}=(\sin\phi,\cos\phi)$.
In term of the elements of ${\sf P}$ we have 
\begin{align}
\sin(2\phi) & =\frac{2{\sf P}_{kn}}{Q}\\
\cos(2\phi) & =\frac{{\sf P}_{nn}-{\sf P}_{kk}}{Q}
\end{align}
where $Q=\sqrt{({\sf P}_{kk}-{\sf P}_{nn})^{2}+4{\sf P}_{kn}^{2}}$ and similarly 
\begin{equation}
 \lambda_{\pm}=({\sf P}_{nn}+{\sf P}_{kk}\pm Q)/2
\end{equation}
We can then quote the final result that generalizes Eq. (\ref{MainRes})
to two control parameters $\theta_n,\theta_k$: 
 \begin{align}
\sigma_{i,{\rm new}}^{2}=\sigma_{i}^{2}\left[1-\frac{S(\rho_{in}^{2}\sigma_{n}^{2}+\rho_{ik}^{2}\sigma_{k}^{2})+2\sigma_{k}^{2}\sigma_{n}^{2}\lambda_{+}\lambda_{-}\Pi-2D\rho_{in}\rho_{ik}\sigma_{n}\sigma_{k}\sin(2\phi)-D\left(\rho_{ik}^{2}\sigma_{k}^{2}-\rho_{in}^{2}\sigma_{n}^{2}\right)\cos(2\phi)}{2\Delta_{nk}\lambda_{+}\lambda_{-}+S\left(\sigma_{n}^{2}+\sigma_{k}^{2}\right)+D\left(\sigma_{n}^{2}-\sigma_{k}^{2}\right)\cos(2\phi)-2D\rho_{nk}\sigma_{n}\sigma_{k}\sin(2\phi)+2}\right]
\end{align}
where 
\begin{align}
& \Pi=\left(\rho_{in}^{2}-2\rho_{ik}\rho_{nk}\rho_{in}+\rho_{ik}^{2}\right)\\
& S=\lambda_{+}+\lambda_{-}\\
& D=\lambda_{+}-\lambda_{-}   
\end{align}
and $\Delta_{nk}=(1-\rho_{nk}^{2})\sigma_{n}^{2}\sigma_{k}^{2}$
is the determinant of the $nk$- submatrix. If $\rho_{nk}=\rho_{in}=\phi=0$
and $\lambda_{-}=p_{k}^{-2}$, we are back to the 1-parameter case.
In the limit of an infinitely strong prior (i.e. knowing the control
parameters precisely) one gets the best possible estimation of the target parameter $\theta_i$
\begin{equation}
\sigma_{i,{\rm new}}^{2} = \sigma_{i}^{2}[1-\frac{\rho_{in}^{2}+\rho_{ik}^{2}-2\rho_{in}\rho_{ik}\rho_{nk}}{1-\rho_{nk}^{2}}]=\sigma_{i}^{2}\frac{\det {\sf R}^{\rm ikn}}{\det {\sf R}^{\rm kn}}
\label{eq:detratio}
\end{equation}
where ${\sf R}^{\sf ikn}$ is the correlation matrix of the three parameters
$\theta_i,\theta_k,\theta_n$, where we use upper indices to avoid confusion with matrix elements which are denoted with downstairs indices.
${\sf R}^{\sf ikn}$ is obtained as in Eq. (\ref{eq:correl}) i.e. by taking only
the $i,k,n$ rows and columns of ${\sf M}^{-1}$ and dividing each entry
on the $a$th row and $b$th column by $\sigma_{a}\sigma_{b}$. Similarly,
${\sf R^{ nk}}$ is the correlation matrix of the parameters $\theta_k$ and $\theta_n$.
It can be shown that the quantity in square brackets lies always between
zero and unity, as it should be since a prior adds information and
the variance cannot increase.

The last expression
Eq. (\ref{eq:detratio}) generalizes to a set of arbitrary many precisely known control parameters in the form 
\begin{equation}
\sigma_{i,{\rm new}}^{2} = \sigma_{i}^{2}\frac{\det {\sf R}^{{\sf i}, \mathcal{C}}}{\det {\sf R}^{\mathcal{C}}}\label{eq:detratio-1}
\end{equation}
where $\mathcal{C}=\{kmn...\}$ represents the set of an arbitrary number of control parameters. Eq.~(\ref{eq:detratio-1}) can be proven as follows.

A well-known generalization of
Cramer's rule for inverting matrices is
\begin{equation}
\det ({\sf A}^{-1})^{J}=\frac{\det ({\sf A})^{J'}}{\det {\sf A}}\label{eq:gencramer}
\end{equation}
where ${\sf A}$ is a $n\times n$ matrix, $\det {\sf A}^{J}$ is the determinant
of the submatrix obtained by keeping only the rows and columns of
the subset $J$, and $J'$ is the complementary subset, such that
every index appears either in $J$ or in $J'$.

Now, since a correlation matrix is connected to the covariance matrix via 
\begin{equation}
 {\sf R} = {\rm diag}(1/\sigma_1, ..., 1/\sigma_n)\ {\sf C}\ {\rm diag} (1/\sigma_1, ..., 1/\sigma_n)
\end{equation}
where ${\rm diag}(1/\sigma_1, ..., 1/\sigma_n)$ is the diagonal matrix of the standard deviations,
Eq. (\ref{eq:detratio-1}) can be
written as
\begin{equation}
\sigma_{i, {\rm new}}^{2}=\frac{\det{\sf C}^{i,\mathcal{C}}}{\det {\sf C}^{\mathcal{C}}}
\label{eq:detratio-1-2}
\end{equation}
We also know that $\sigma_{i,{\rm new}}^{2}$ is obtained by maximizing the control parameters, i.e. by inverting the covariance matrix
${\sf C}$, striking out the rows/columns corresponding to the $\mathcal{C}$ subset
so to produce the matrix $({\sf C}^{-1})^{\mathcal{C}'}$, where $\mathcal{C}'$ is the complementary set of parameters to $\mathcal{C}$. This matrix then needs to be inverted again and
the entry ${\sf C}_{ii}$ is then $\sigma_{i,{\rm new}}^{2}$. 

This entry will however be given by the minor determinant
of the $ii$-element of $({\sf C}^{-1})^{\mathcal{C}'}$ divided by the determinant
of $({\sf C}^{-1})^{\mathcal{C}'}$ That is, using in the last step Eq. (\ref{eq:gencramer}),
\begin{equation}
\sigma_{i, {\rm new}}^{2}=\frac{\det ({\sf C}^{-1})^{(i,\mathcal{C})'}}{\det ({\sf C}^{-1})^{\mathcal{C}'}}=\frac{\det {\sf C}^{i,\mathcal{C}}}{\det {\sf C}^{\mathcal{C}}}
\label{eq:detratio-1-2-1}
\end{equation}
where the notation $(i,\mathcal{C})'$ means the index subset formed by all
the indexes except the $i$-th one and the $\mathcal{C}$ subset.

Eq. (\ref{eq:detratio-1}) then gives a handy expression to determine the best possible estimate
of the target parameter in the limit of an exact determination of
the control parameters. Most experiments provide directly the parameter
covariance or correlation matrix, obtained for instance through a
MonteCarlo Markov Chain procedure; our expression can be immediately
applied to the correlation matrix and no matrix inversion is needed.

Finally, we quote for completeness a  further generalization to any number of target parameters but valid only
when {\it all} the other parameters are considered control parameters, and again in the limit of an infinitely
strong prior. This relation is well known in statistics because it gives the {\it conditional covariance}, i.e. the covariance
when some of the random variables are known.
Denoting with $\mathcal{T}$ the set of target parameters, we have
\begin{equation}
\det {\sf C}^{\mathcal{T}}_{(new)}=\frac{\det {\sf C}^{\mathcal{T},\mathcal{C}}}{\det {\sf C}^{\mathcal{C}}}\label{eq:detratio-1-1}
\end{equation}
The inverse of the determinant of the submatrix ${\sf C}^{\mathcal{T}}$ is often
called a figure-of-merit. So this expression tells us how much the
figure-of-merit improves when all the other  parameters are fixed, i.e.  the best possible figure-of-merit one can achieve
by improving the constraints on correlated parameters. 

\bibliography{sherman2Notes}
\bibliographystyle{mn2e}

\end{document}